# Quantum de Sitter Space-Time and the Dark Energy

Liao Liu

Department of Physics，Beijing Normal University
Beijing 100875，P. R. China
Liuliao1928@yahoo.com.cn

## Abstract

Three years ago, we introduced a new way to quantize the static Schwarzschild black hole(SSBH), there the SSBH was first treated as a single periodic Euclidean system and then the Bohr-Sommerfeld quantum condition of action was used to obtain a quantum theory of Schwarzschild black hole.[1] Now in this short report, we try to extend the above method to quantize the static de Sitter(SDS) space-time and establish a quantum theory of both SDS space and dark energy.

The metric of Lorentzian SDS space-time reads

$$ds^2 = -(1 - \frac{1}{3}\Lambda r^2)dt^2 + (1 - \frac{1}{3}\Lambda r^2)^{-1}dr^2 + r^2 d\Omega_2^2 \tag{1}$$

or

$$ds^2 = -(1 - H^2 r^2)dt^2 + (1 - H^2 r^2)^{-1}dr^2 + r^2 d\Omega_2^2 \quad (H = \sqrt{\Lambda/3}) \tag{2}$$

It is known there is a coordinate singularity at $r = H^{-1}$, which is just the position of event horizon of SDS. This coordinate singularity can be get ride by introducing a Kruskal-like coordinate as follows [2]

$$ds^2 = \frac{1}{H^2}(uv - 1)^{-2}[-4du\,dv + (uv + 1)^2 d\Omega_2^2] \tag{3}$$

where $\quad r = (\frac{uv+1}{1-uv})\sqrt{\frac{3}{\Lambda}} = (\frac{uv+1}{1-uv})H^{-1}$

$$\exp(2Ht) = -v u^{-1} \tag{4a}$$

or $\quad u = e^{-H^{-1}t}(\frac{2}{H^{-1}r+1} - 1)^{1/2}$

$$v = e^{H^{-1}t}(\frac{2}{H^{-1}r+1} - 1)^{1/2} \tag{4b}$$

We see from equation (4 a, b) that the Kruskal-like Euclidean manifold has an imaginary period of $T = 2\pi H$. So a very important thing now we have is that the Euclidean Kruskal-like SDS manifold can be looked upon as a single period $T = 2\pi H$ system. Now let us try to apply the so-called Bohr-Sommerfeld action quantization principle to the Euclidean Kruskal-like de Sitter manifold.

As is known, the action I , action variable $I_v = \oint pdq$ and the Hamiltonian $\mathcal{H}$ of any single periodic system should satisfied the relation [3]

$$I = I_v - \oint \mathcal{H} dt \tag{5}$$

and the Hamiltonian constraint

$$\mathcal{H}(h_{ij}, \pi_{ij}) = 0$$

So equation (5) becomes

$$I = I_v = \oint p\,dq \tag{6}$$

Now according to the Bohr-Sommerfeld action quantization condition, we have

$$I_v = \oint pdq = 2\pi(n+\frac{1}{2})$$

or

$$I = I_v = 2\pi(n+\frac{1}{2}) \tag{7}$$

Let's remember the contribution to $I$ comes from two terms, i.e. the one is given by volume integral

$$-\frac{1}{16\pi}\int_M d^4x\sqrt{g}\,(R-2\Lambda) \tag{8}$$

the other is Gibbons-Hawking surface term

$$-\frac{1}{4\pi}\int_{\partial M} k\sqrt{h}\,d^3x \tag{9}$$

Calculation from expressions (8) and (9) gives respectively[4]

$$-\frac{1}{16\pi}\int_M d^4x\sqrt{g}\,(R-2\Lambda) = -\frac{\pi}{H^2} \tag{10}$$

and

$$-\frac{1}{4\pi}\int_{\partial M} k\sqrt{h}\,d^3x = \frac{2\pi}{H^2} \tag{11}$$

So the action $I$ of Euclidean Kruskal-like SDS  manifold is

$$I = \frac{\pi}{H^2} \tag{12}$$

Expression (7) and (12) gives

$$H_n^{-2} = 2n+1$$

or

$$\Lambda_n^{-1} = \frac{1}{3}(2n+1) \tag{13}$$

Thus the Bohr-Sommerfeld action principle forced the cosmology constant $\Lambda$ or SDS to take discrete value as shown by Expression (13). Now, if we define the dark energy density inside the DS space by [4]

$$\rho_\Lambda = \frac{\Lambda_n}{8\pi G} \tag{14}$$

then the total dark energy inside the cosmological event horizon of certain SDS universe should as follows[4]

$$E_{\Lambda_n} = \frac{1}{2} H_n^{-1} = \frac{1}{2}\sqrt{2n+1}\, m_p \tag{15}$$

The ground state dark energy of any SDS is then

$$E_{n=0} = \frac{1}{2} m_p$$

Their temperature is

$$T = \frac{H_0}{2\pi} = \frac{l_p^{-1}}{2\pi} = \frac{1}{2\pi}(G\hbar c^{-3})^{-\frac{1}{2}} = \frac{1}{2\pi}\times[1.6\times 10^{-33}]^{-1} \simeq 10^{33}\, \text{deg}.$$

which is the temperature of dark energy in the ground state of SDS space. According to Everett's interpretation of quantum mechanics, the quantum cosmology may lead to a multiplicity of realities. So the cosmology wave function will provide many $O_4$ bubbles with different cosmological constant $\Lambda$, which should be quantized in our quantization scheme as given by (15). It seems a foam-like sea of quantum $O_4$ bubbles or network of $O_4$ topology. may existed in the very early universe. Their ground state as a whole will form a very high temperature background of dark energy in our very early universe.

At last, we arrive at the following conclusion. The dark matter may be just the relics of primordial small black hole of life-time not longer than that of our universe [5]. The dark energy may be just the relics or ground state of quantum $O_4$ bubbles in the very early universe.

## Acknowledgement

I would like to give my sincere thanks to Prof. C. G. Huang for his comments.

# Addenda

Though nowadays no prevalent quantum gravity theories exists, many attempts to quantize black holes and de Sitter space are still underway. A collection of some important relevant papers are included in the addenda.

[1]  J.D.Bekenstein, *Lett. Nuovo Cimento* 11(1974)487
[2]  J.D.Bekenstein,V.F.Mukhanov, *Phys.Lett.B*360(1995)7
[3]  J.D.Bekenstein, *gr-qc*/9710076
[4]  J.D. Bekenstein, G. Gour, *Phys.Rev.D*66(2002)024005
[5]  L.Liu. S.Y. Pei, *Chinese Phys.Lett.* 21(2004)1887
[6]  L.Liu. S.Y.Pei, *Acta Physica Sinica* 55(2006)4980
[7]  L.Xiang, Y.G.Shen, *Phys.Lett.B*602(2004)226
[8]  T.Padmanabhan, *Class. Quantum Grav.*19(2002)L167